\newcommand\beq{\begin{eqnarray}}
\newcommand\eeq{\end{eqnarray}}
\newcommand\meff{m_{\rm eff}}
\newcommand\missET{E_T^{\rm miss}}

\def\lsim{\mathrel{\rlap{\lower4pt\hbox{$\sim$}}
    \raise1pt\hbox{$<$}}}                
\def\gsim{\mathrel{\rlap{\lower4pt\hbox{$\sim$}}
    \raise1pt\hbox{$>$}}}

\documentclass[
amsmath,
prd,nofootinbib,floatfix,11pt 
]{revtex4}

\allowdisplaybreaks
\usepackage{graphicx}
\usepackage{setspace}

\begin{document}
\renewcommand{\theequation}{\arabic{section}.\arabic{equation}}
\begin{flushright}
ANL-HEP-PR-11-77
\end{flushright}

\title{\Large%
\baselineskip=21pt
Compressed supersymmetry after 1 fb$^{-1}$ at the
Large Hadron Collider}

\author{Thomas J.~LeCompte$^1$ and Stephen P. Martin$^{2,3,4}$}
\affiliation{
{\it 1) Argonne National Laboratory, Argonne IL 60439, USA} 
\\
{\it 2) Department of Physics, Northern Illinois University, 
DeKalb IL 60115, USA} 
\\
{\it 3) Fermi National Accelerator Laboratory, P.O. Box 500, 
Batavia IL 60510, USA}
\\
{\it 4) \mbox{Kavli Institute for Theoretical Physics, University of 
California, 
Santa Barbara CA 93106, USA}}
}

\begin{abstract}\normalsize \baselineskip=15pt 
We study the reach of the Large Hadron Collider with 1 fb$^{-1}$ of 
data at $\sqrt{s} = 7$ TeV for several classes of supersymmetric models 
with compressed mass spectra, using jets and missing transverse 
energy cuts like those employed by ATLAS for Summer 2011 data.
In the limit of extreme compression, the best limits come from signal 
regions that do not require more than 2 or 3 jets and that remove backgrounds 
by requiring more missing energy rather than higher effective mass.
\end{abstract}


\maketitle

\tableofcontents

\vfill\eject
\baselineskip=16pt

\setcounter{footnote}{1}
\setcounter{page}{2}
\setcounter{figure}{0}
\setcounter{table}{0}

\section{Introduction}
\label{sec:intro}
\setcounter{equation}{0}
\setcounter{footnote}{1}

The Large Hadron Collider (LHC) is now testing the proposal that 
supersymmetry \cite{SUSYreviews} (SUSY) is the solution to the hierarchy 
problem associated with the electroweak scale. At this writing, there 
have been no hints of supersymmetry, defying expectations 
based on the sensitivity of the Higgs potential to superpartner masses in 
many models, including the popular ``mSUGRA" (minimal supergravity) scenario. It is 
possible that the failure of SUSY to appear is simply due to the up and 
down squarks being very heavy, as their production otherwise gives 
the 
strongest bounds. 
Another possibility is that the superpartners are not so heavy, but are 
difficult to detect because of a compressed mass spectrum, leading to 
much smaller visible energy than in mSUGRA benchmark cases. For our 
purposes here, compressed SUSY refers to the situation in which the mass 
ratio between the lightest supersymmetric particle (LSP) and the gluino 
is significantly smaller than the prediction $m_{\rm 
LSP}/m_{\tilde g} \sim 1/6$ of mSUGRA.

In a previous paper \cite{LeCompte:2011cn}, we investigated the reach of 
2010 data from ATLAS, consisting of 35 pb$^{-1}$ of collisions at 
$\sqrt{s}=7$ TeV, for several classes of compressed SUSY models. In the 
present work, we will update this analysis to correspond to the 1.04 
fb$^{-1}$ data set analyzed by ATLAS in ref.~\cite{ATLASsummer2011}. Not 
only does this represent a huge increase in integrated luminosity, but 
also a revised set of signal regions compared to the 2010 data set 
analyses. The ATLAS analysis presents exclusion results for mSUGRA 
models, 
and for simplified models containing only squarks and a gluino but with 
the LSP mass held fixed at 0. In both cases the models tested are very 
far from the compressed case in which the mass difference between the 
gluino and the LSP is smaller. Our aims here are to see how the 
exclusions 
found for mSUGRA and simplified models 
translate into exclusions on compressed SUSY, for the various ATLAS 
signal regions, with particular attention to the exclusions that can be 
made in the most difficult case of very high compression. Other recent 
studies 
that include compressed SUSY and other non-mSUGRA searches at the LHC 
in a similar spirit can be found in 
\cite{Baer:2007uz}-\cite{Papucci:2011wy}.

The rest of this paper is organized as follows. Section 
\ref{sec:procedures} describes the signal regions and our procedures. 
Section 
\ref{sec:models} describes four classes of models, each of which
depends on two parameters, the gluino mass 
$M_{\tilde g}$ and a 
compression parameter $c$,  
which are independently and  continuously dialed to vary the overall 
superpartner mass scale and the ratio of 
gluino to LSP 
masses. Section \ref{sec:results} gives results for the acceptances 
for these models with the various signal regions, and estimated 
exclusions based on the 1.04 fb$^{-1}$ data set. Section \ref{sec:conclusion}
contains some concluding remarks.

\section{Procedures and signal requirements}
\label{sec:procedures}
\setcounter{equation}{0}
\setcounter{footnote}{1}

For this paper, we used the same tools as in our earlier work 
\cite{LeCompte:2011cn}.
{\tt 
MadGraph/MadEvent 4.4.62} \cite{MGME} was used to 
generate hard scattering events using CTEQ6L1 \cite{CTEQ}
parton distribution functions, {\tt Pythia 6.422} \cite{Pythia} for 
decays and showering and hadronization, and PGS 4 \cite{PGS} for detector 
simulation. In SUSY models with compressed mass spectra, it is 
important to correctly generate jets beyond the hard scattering event, by
matching correctly (without overcounting) between matrix-element and 
showering/hadronization 
software generation of additional jets. We did this by generating each 
lowest-order process together with the same process with one additional 
jet at the matrix-element level, followed by MLM matching with 
$P_T$-ordered showers with the shower-$K_T$ scheme with $Q_{\rm cut} = 
100$ GeV, as described in \cite{matching} and implemented in the 
MadGraph/MadEvent package. (It is much more time-consuming to include 
up to two 
extra jets at the matrix-element level. 
We found with some sample testing that 
it did not make a significant difference with 
our setup even for very compressed 
superpartner mass spectra.) For 
the detector simulation, we used the default ATLAS-like parameter 
card file provided with the PGS distribution, but with a jet 
cone size of $\Delta R = 0.4$. Cross-sections were normalized to the 
next-to-leading order output of 
{\tt Prospino 2.1} \cite{prospino}. 

To define signals, we follow (a slightly simplified version of) the ATLAS 
cuts for multijets+$\missET$ from ref.~\cite{ATLASsummer2011}.
The signal requirements are summarized in Table \ref{tab:cuts}. 
\begin{table}
\begin{tabular}[c]{lccccc}
& A & B & C & D & E 
\\
\hline
\hline
Leading jet $p_T$ [GeV] & $>130$ & $>130$ & $>130$ & $>130$ & $>130$
\\
Number of jets $n$ & $\geq 2$ & $\geq 3$ & $\geq 4$ & $\geq 4$ & $\geq 4$ 
\\ 
$p_T(j_n)$ [GeV] & $>40$ & $>40$ & $>40$ & $>40$& $>80$
\\
$\meff$ [GeV] & 
\phantom{x}$>1000$\phantom{x} & 
\phantom{x}$>1000$\phantom{x} & 
\phantom{x}$>500$\phantom{x} & 
\phantom{x}$>1000$\phantom{x} & 
\phantom{x}$>1100$\phantom{x} \\
$\missET/\meff$ & $>0.3$ & $>0.25$ & $>0.25$ & $>0.25$ & $>0.2$
\\
\hline
ATLAS $\sigma\times{\rm Acc}$ [fb]  
\phantom{xx}& $<22$ & $<25$ & $<429$ & $<27$ & $<17$
\\
\end{tabular}
\caption{\label{tab:cuts} Summary of cuts for the signals A, B, C, D, E 
simulated here, following the ATLAS 2011 data analyses for 1.04 fb$^{-1}$ 
\cite{ATLASsummer2011}. Also shown on the last line are the ATLAS 95\% CL 
bounds on the non-Standard Model contribution to the cross-section times 
acceptance in the five signal regions. (In the case of signal region E 
only, the $\meff > 1100$ GeV requirement involves a sum over all jets 
with $p_T > 40$ GeV, but the $\meff$ used in the $\missET/\meff$ cut is a 
sum over only the leading 4 jets.)}
\end{table}
Events are required to have at least one jet with $p_T>130$ GeV.
The signal regions A, B, C, and D also require at least 2, 3, 4, and 4
jets with $p_T > 40$ GeV, respectively, while signal region $E$ requires 
at least 4 jets with $p_T>80$ GeV. 
These jets must have $|\eta| < 2.5$. 
The leading three jets are required to 
be isolated
from the missing transverse momentum according to 
$\Delta \phi(\vec{p}_T^{\phantom{.}\rm miss},j) > 0.4$. 
The effective mass 
$\meff$ is defined as the scalar sum of the $\missET$ and the $p_T$'s of: 
the leading 2 jets for A; the leading 3 jets for B, the leading 4 jets 
for C, D; and all jets with $p_T > 40$ GeV for E. 
Then $\meff$ is required to exceed 1000, 1000, 500, 1000, and 1100 GeV 
respectively for signal regions A, B, C, D, E. In addition, a cut is 
imposed on the ratio $\missET/\meff$ of 0.3, 0.25, 0.25, 0.25, and 0.2
for A, B, C, D, E respectively. (For signal region E, only the 4 leading 
jet
$p_T$'s are summed over in the $\meff$ used in the $\missET/\meff$ cut,
even though the $\meff$ cut uses an inclusive sum over jets.)
Note 
that these cuts automatically imply a lower limit on 
$\missET$ of 300, 250, 125, 250 GeV for signals A, B, C, D,
respectively. For signal region E, a cut $\missET > 130$ GeV is imposed,
although on an event-by-event basis this is usually superseded by the 
$\missET/\meff$ cut. There is a veto of events with
leptons $\ell = (e,\mu)$ with $|\eta| < 2.4$ and  
(2.47) for muons (electrons), and $p_T^\ell > 20$ GeV that are farther 
than $\Delta R = 
\sqrt{(\Delta \eta)^2 + (\Delta \phi)^2} > 0.4$ from the nearest jet. 
Also shown on the last line of Table \ref{tab:cuts} are the ATLAS 95\% CL 
limits on non-Standard-Model contributions to the signal regions after acceptance and efficiency, as reported in ref.~\cite{ATLASsummer2011}. These will be used below to estimate the reach for compressed SUSY models.
ATLAS also has searches requiring higher jet multiplicities \cite{Aad:2011qa},
leptons \cite{Aad:2011iu}, and $b$ tagging \cite{ATLASb}, but these searches
give significantly less reach for the
compressed SUSY models we consider.
Comparable searches by CMS have been reported in \cite{Chatrchyan:2011zy}, 
\cite{CMSleptons}, and \cite{CMSb}; we choose to use the ATLAS results 
only for reasons of convenience and familiarity.

Because our tools for generating SUSY signal events and simulating 
detector response are not the same as those used by ATLAS, the 
cross-section and acceptance results found below clearly cannot be 
interpreted in exact correspondence to the ATLAS ones. However, we have 
checked that the results of our analysis methods correlate well to those 
in ref.~\cite{ATLASsummer2011} for a sample of mSUGRA 
models used there. For mSUGRA models with $\tan\beta=10$, $A_0=0$, $m_0 = 
100$ GeV, and $m_{1/2} = 180$, 210, 240, 270, 300, 330, 360, 390, 420, 
450, and 480 GeV, we found agreement with the ATLAS 
acceptance$*$efficiency to be typically better than 15\%, while for the 
same parameters but $m_0 =660$ GeV, the agreement was usually at the 30\% 
level or better. Keeping these inevitable 
differences in mind, at least 
an 
approximate estimate of the true detector response may still be gleaned 
from the results below, and the general trends should be robust.

\section{Compressed SUSY models}
\label{sec:models}
\setcounter{equation}{0}
\setcounter{footnote}{1}

In this section, we define the compressed SUSY models for our study. 
Following our earlier work ref.~\cite{LeCompte:2011cn}, 
let us parameterize the electroweak gaugino masses at the TeV scale in 
terms of the gluino physical mass as
\beq
M_1 \>=\> \left (\frac{1 + 5 c}{6}\right )M_{\tilde g},
\qquad\qquad
M_2 \>=\> \left (\frac{1 + 2 c}{3}\right )M_{\tilde g}.
\label{eq:cmodelgauginomasses}
\eeq
Here $c$ parameterizes the degree of compression. 
The value $c=0$ gives an 
mSUGRA-like mass spectrum with gaugino masses approximately equal at 
$M_{\rm GUT} = 2.5 \times 10^{16}$ GeV, 
and $c = 1$ gives a completely compressed spectrum in 
which the gluino, wino, and bino masses are equal at the TeV  
scale. The gluino mass $M_{\tilde g}$ is treated as the independent 
variable input 
parameter that sets the superpartner mass scale.
We also select $\tan\beta = 10$ and positive $\mu = M_{\tilde g} + 200$ GeV 
to compute the physical masses of charginos $\tilde C_i$ 
and neutralinos $\tilde N_i$.
For our first class of models, we take the first- and second-family 
squark masses to be:
\beq
m_{\tilde u_R} = m_{\tilde d_R} = m_{\tilde u_L} = 0.96 M_{\tilde g},
\qquad
m^2_{\tilde d_L} = m_{\tilde u_L}^2 - \cos(2\beta) m_W^2,
\eeq
and sleptons are taken degenerate with the squarks (so too heavy to 
appear in 
chargino and neutralino two-body decays). The top squark masses are
taken to be $m_{\tilde t_1} = M_{\tilde g} - 160 + c (180 - 0.09 
M_{\tilde g})$ and $m_{\tilde t_2} = M_{\tilde g} + 25$, in GeV.
The lightest Higgs mass is fixed at $m_{h^0} = 115$ GeV, and the heavier 
Higgs masses with $m_{A^0} = 0.96 M_{\tilde g}$.
These choices provide relatively smoothly varying 
branching ratios as the compression parameter $c$ is varied, although
transitions of $\tilde N_2$ and $\tilde C_1$ decays from on-shell 
to off-shell weak bosons are 
inevitable as the compression increases. 
In particular, the reason for the choice for the
parameterization of the stop masses is to avoid suddenly
turning on or off any 2-body decay modes as the parameter c is varied
within each model line, by making sure that the the
gluinos cannot decay to stops by kinematics for any of these models. 
The choices 
for $\tan\beta$ and $\mu$ are arbitrary,
and not very much would change if they were modified (within
some reasonable range). 
We refer to the two-parameter class of models spanned by varying 
$(M_{\tilde g}, c)$ as models I.

\begin{figure}[!tbp]
\begin{minipage}[]{8.3cm}
\includegraphics[width=8.3cm,angle=0]{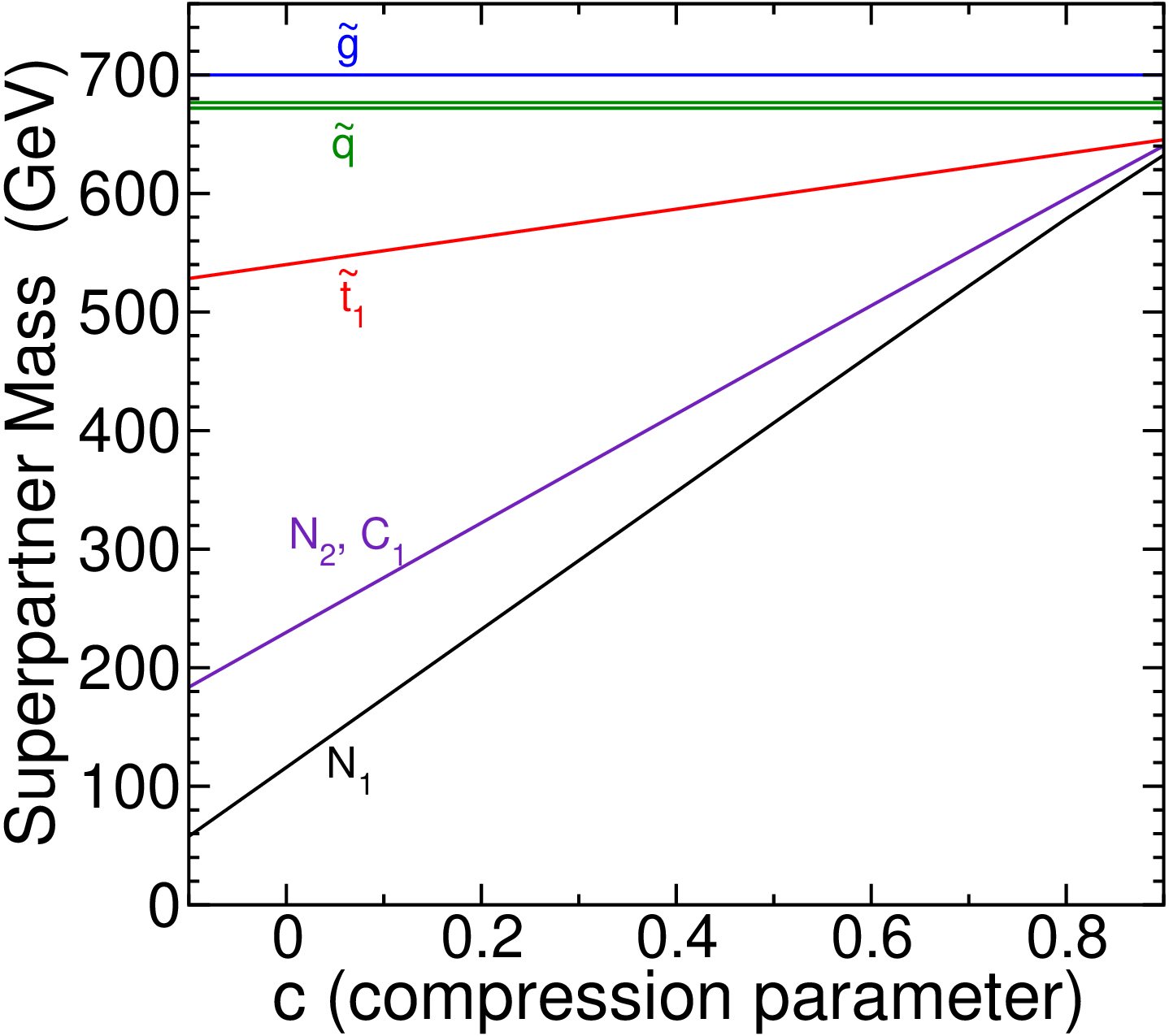}
\end{minipage}
\hspace{0.6cm}
\begin{minipage}[]{4.4cm}
\caption{\label{fig:cmasses}
The masses of the most relevant superpartners for the models of class I
defined in section \ref{sec:models}, as a function of the 
compression 
parameter $c$, for fixed $M_{\tilde g} = 700$ GeV. The case $c=0$ 
corresponds to an mSUGRA-like model.}
\end{minipage}
\end{figure}

The masses of the most 
relevant superpartners are shown in Figure \ref{fig:cmasses} for the case 
$M_{\tilde g} = 700$ GeV, 
illustrating the effect of the compression parameter $c$ on the spectrum.
In this class of models, gluino and squark production dominate at the 
LHC. The gluino decays mostly by the two-body mode $\tilde g \rightarrow 
\overline q \tilde q$ or $q\tilde{\overline q}$, and right-handed squarks 
decay mostly directly to the LSP, $\tilde q_R \rightarrow q \tilde N_1$, 
while left-handed squarks decay mostly to wino-like charginos and 
neutralinos, $\tilde q_L \rightarrow q' \tilde C_1$ and $q \tilde N_2$. 
The latter decay through on-shell or off-shell weak bosons: $\tilde C_1 
\rightarrow W^{(*)} \tilde N_1$ and $\tilde N_2 \rightarrow Z^{(*)} 
\tilde N_1$, or $\tilde N_2 \rightarrow h \tilde N_1$ when it is 
kinematically allowed. The visible energy in each event from these decays 
clearly decreases as the compression factor $c$ increases, because of the 
reduction in available kinematic phase space.

We define a second class of ``heavy squark" models, II, which are the 
same as above but 
with all squarks taken very heavy, $M_{\tilde Q} = M_{\tilde g} + 1000$ 
GeV. Thus when $c=0$, the model classes I and II correspond 
approximately to 
mSUGRA with small and large $m_0$, respectively. 
In these heavy squark models, the most important production cross-section 
is from 
gluino pair production, with subsequent gluino decays $\tilde g 
\rightarrow \tilde C_1 q \bar q'$ and $\tilde N_2 q \bar q$ and $\tilde 
N_1 q \bar q$, with the first two typically dominating. The wino-like 
states then decay through on-shell or off-shell weak bosons, 
depending on the mass difference from the compression: 
$\tilde C_1 \rightarrow W^{(*)} \tilde N_1$ and $\tilde N_2 \rightarrow 
Z^{(*)} 
\tilde N_1$ or $h\tilde N_1$, with the last dominating if 
kinematically allowed. 

One motivation for compressed SUSY is that taking $M_3/M_2$ much less 
than the 
mSUGRA prediction can significantly ameliorate 
\cite{KaneKing,compressedSUSYa,compressedSUSYc} the supersymmetric little 
hierarchy problem. This provides motivation to consider compressed SUSY 
models in which winos are heavier than the gluino. Therefore, we define a 
third and fourth class of models, III (``heavy winos") and 
IV (``heavy
winos and squarks"), to be the 
same as models I 
and II respectively, but with $M_2 = M_{\tilde g} + 100$ GeV in each 
case. 
In models III, all first- and 
second-family squarks decay directly to the LSP: $\tilde q \rightarrow q 
\tilde N_1$, while in models IV the squarks decouple from the discovery 
or best exclusion limit processes. The gluino has direct two-body decays 
to quarks and squarks as before.

The model classes I, II, and III were exactly those we used in 
ref.~\cite{LeCompte:2011cn} in the context of limits obtainable with 35 
pb$^{-1}$ at LHC, while the model class IV corresponds approximately to 
the heavy squark limit of the simplified gluino/squark models in 
\cite{ATLASsummer2011}, but with neutralino LSP masses that are here 
non-zero and vary continuously with $c$. Thus the models that we discuss 
here provide a quite different slicing through the MSSM parameter space 
than those found in the experimental collaboration papers. We now proceed 
to use them to examine how the ATLAS exclusions on non-Standard Model 
cross-section times acceptance times efficiency impact the parameter 
space for SUSY with compressed mass spectra.

\section{Results of simulations}
\label{sec:results}
\setcounter{equation}{0}
\setcounter{footnote}{1}

\begin{figure}[!tbp]
\mbox{\includegraphics[width=8.3cm,angle=0]{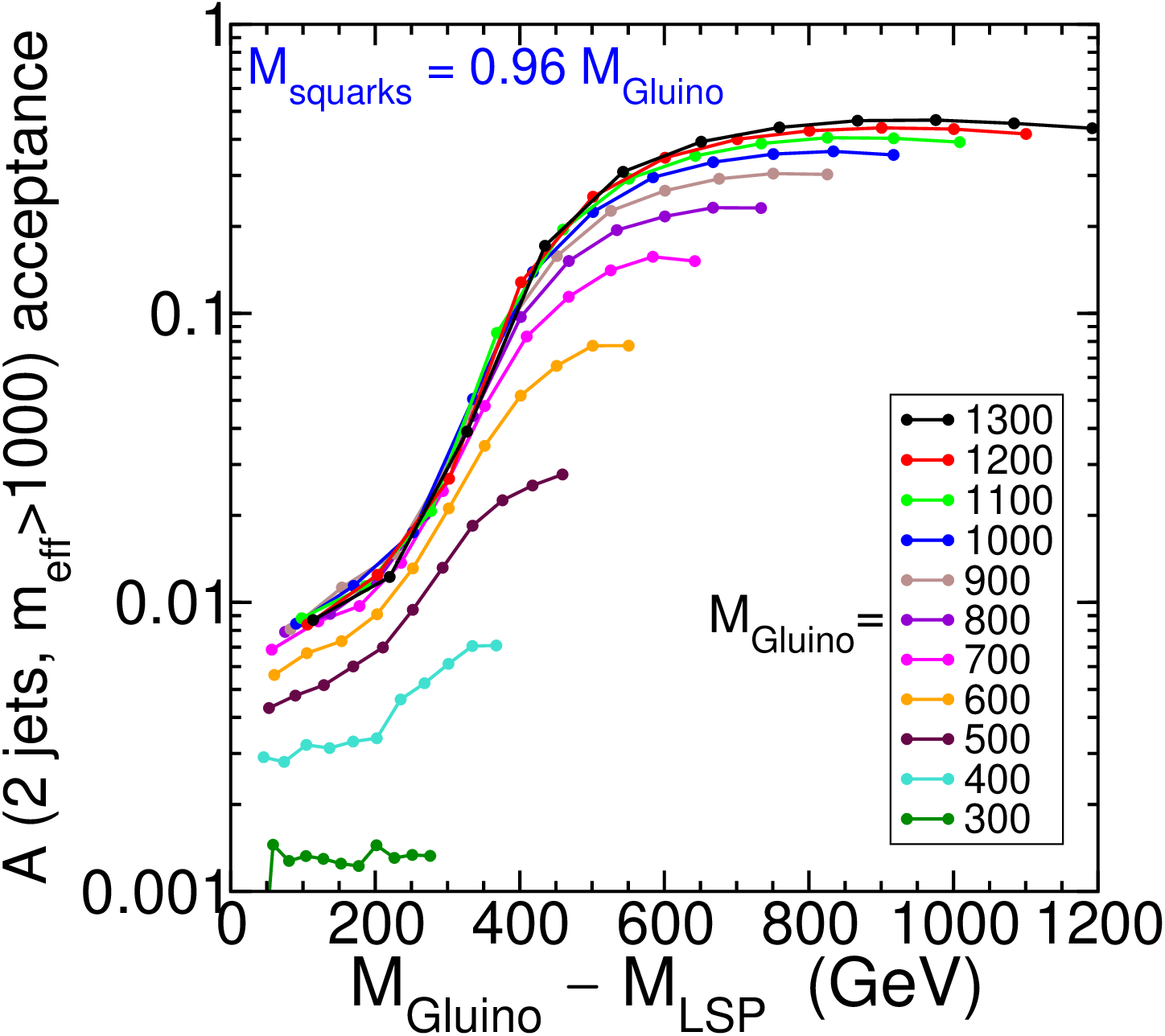}
\includegraphics[width=8.3cm,angle=0]{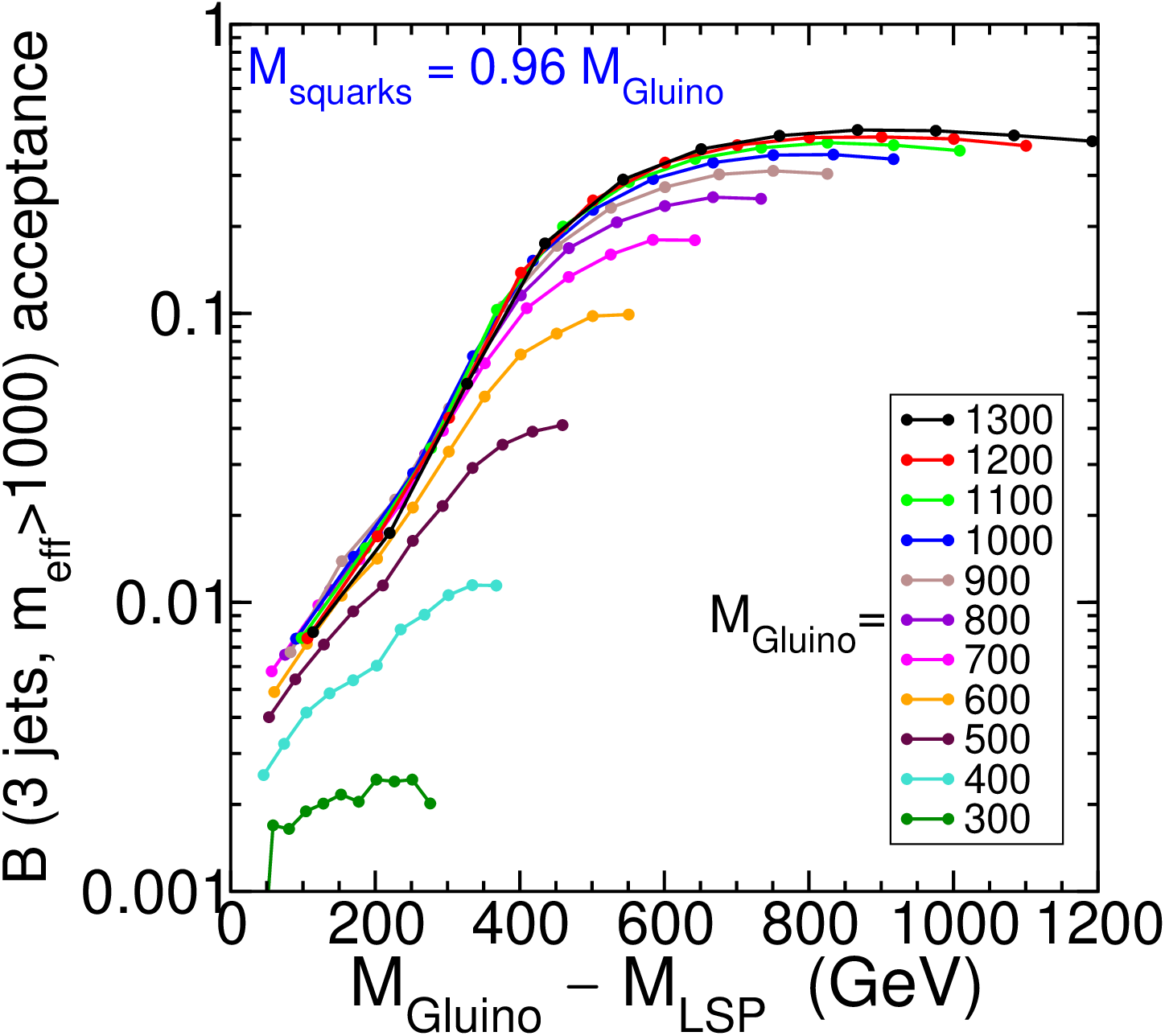}}

\vspace{0.2in}

\mbox{\includegraphics[width=8.3cm,angle=0]{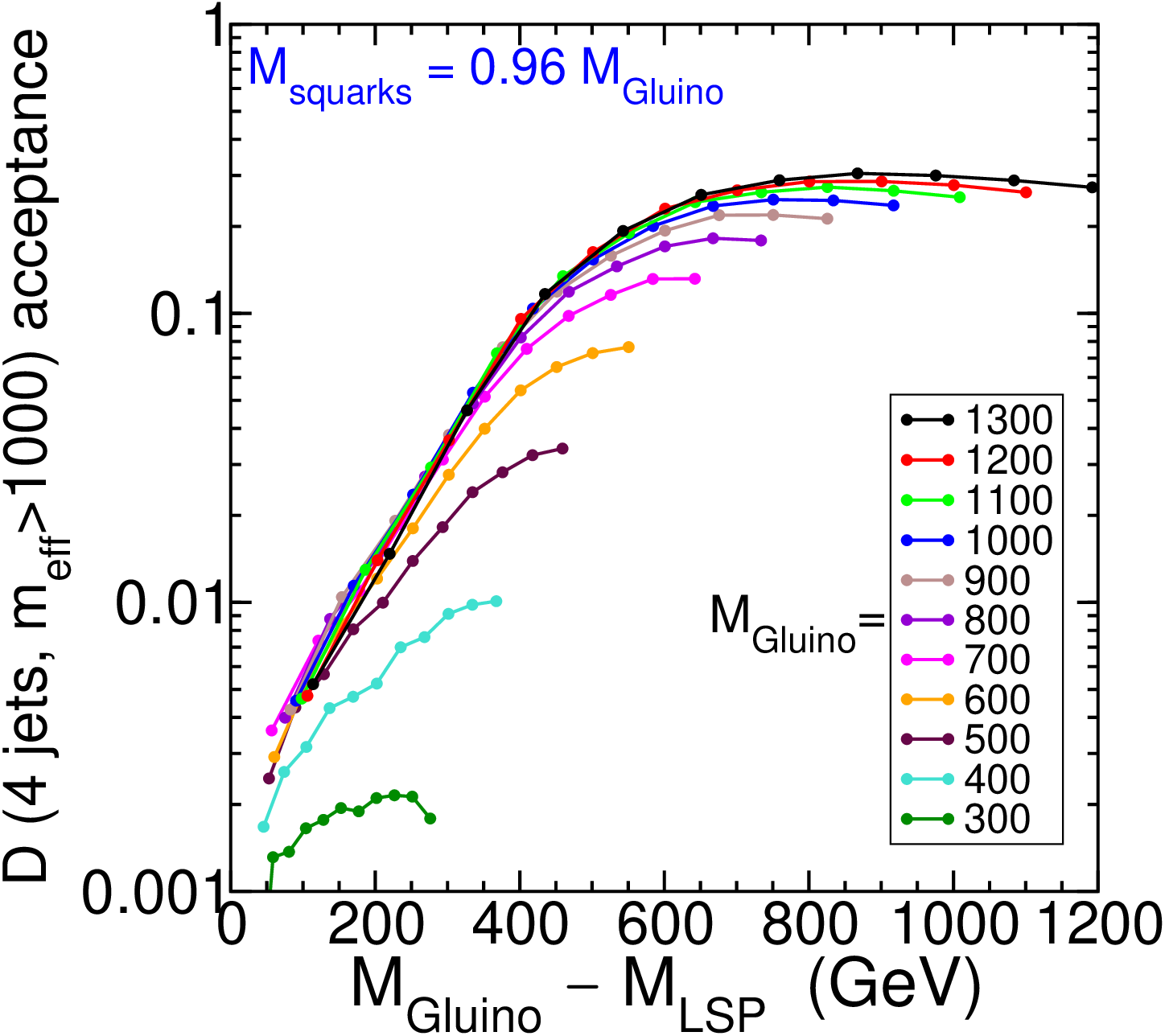}
\includegraphics[width=8.3cm,angle=0]{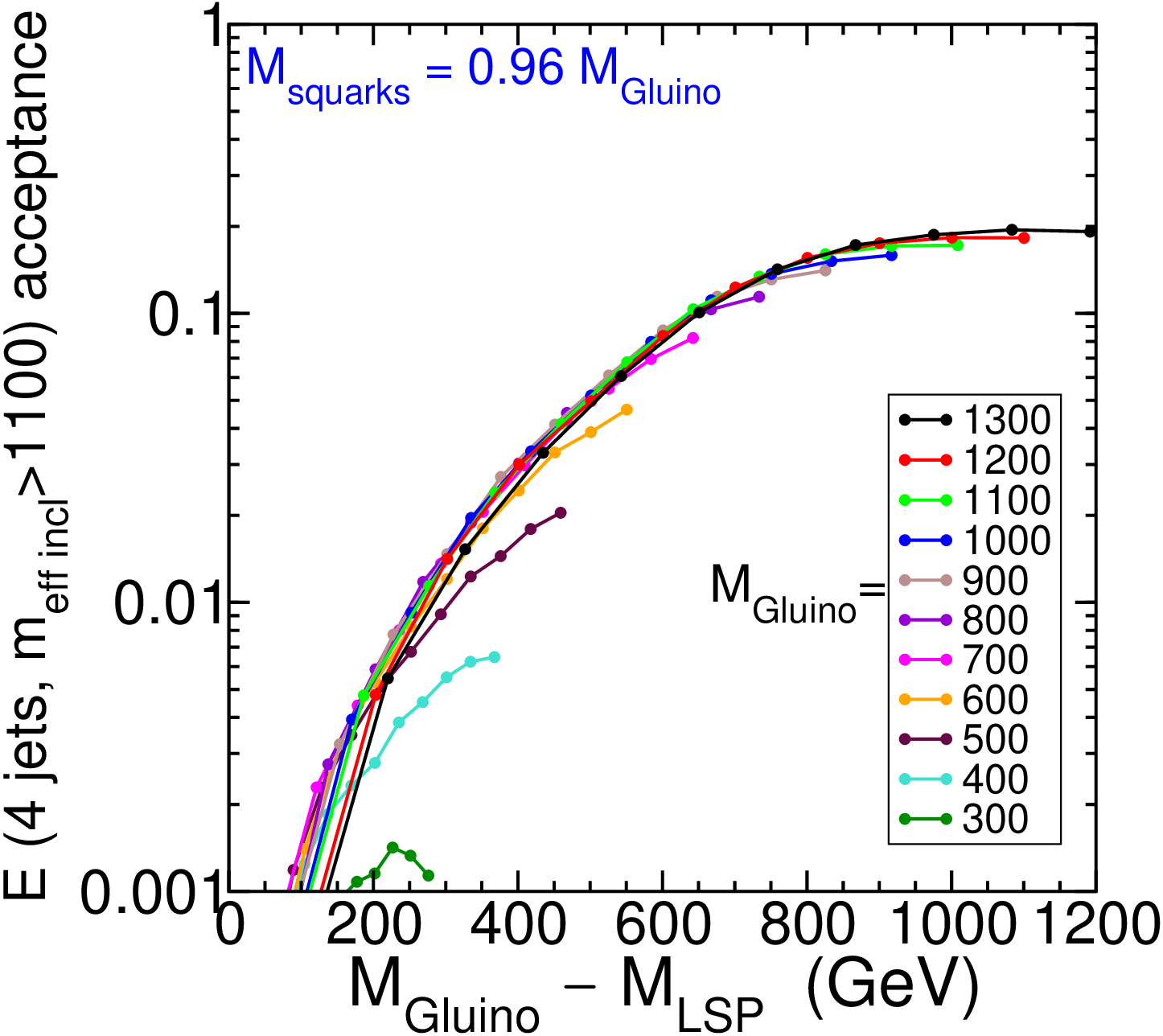}}
\caption{\label{fig:acc}
Acceptances 
for the four signal regions A, B, D, E defined in Table \ref{tab:cuts},
for the class of models I defined in the text.
The lines on each graph correspond to different values of the gluino 
mass. The dots on each line correspond to compression factors $c = -0.1, 0, 0.1, \ldots 0.9$
from right to left.}
\end{figure}

In compressed SUSY models, the visible energy in jets is reduced compared 
to mSUGRA models, leading to a significant reduction in acceptance for
signal events. In Figure \ref{fig:acc}, we show the fractional acceptances
after all cuts for the models in class I with $c = -0.1, 0, 0.1, \ldots, 0.9$
and $M_{\tilde g} = 300, 400, \ldots, 1300$ GeV. Results are shown for each of
the signal regions A, B, D, and E. (We find that signal region C is not competitive for
setting limits in any of the models we consider, see Figure \ref{fig:sigacc} below, 
so it is not included in Figure 
\ref{fig:acc}.)

For fixed values of the mass difference $M_{\tilde g} - M_{\rm LSP}$, the 
acceptance tends to approach a limit for sufficiently large $M_{\tilde 
g}$. Conversely, for sufficiently large $M_{\tilde g} - M_{\rm LSP}$, the 
acceptance tends to be relatively flat, but falls dramatically for 
$M_{\tilde g} - M_{\rm LSP} \lsim 450$ GeV for signal regions A, B, 
and D, 
and for an even larger range of the mass difference for the high-mass 
signal region E. For severe compression $M_{\tilde g} - M_{\rm 
LSP} < 150$ GeV, the signal regions A and B can be seen to retain 
acceptance more than the signal regions D and E do, although in each 
case the acceptance declines to well below 1\% for the most extreme 
compression, even when the gluino mass is very large.

\begin{figure}[!tbp]
\mbox{\includegraphics[width=8.3cm,angle=0]{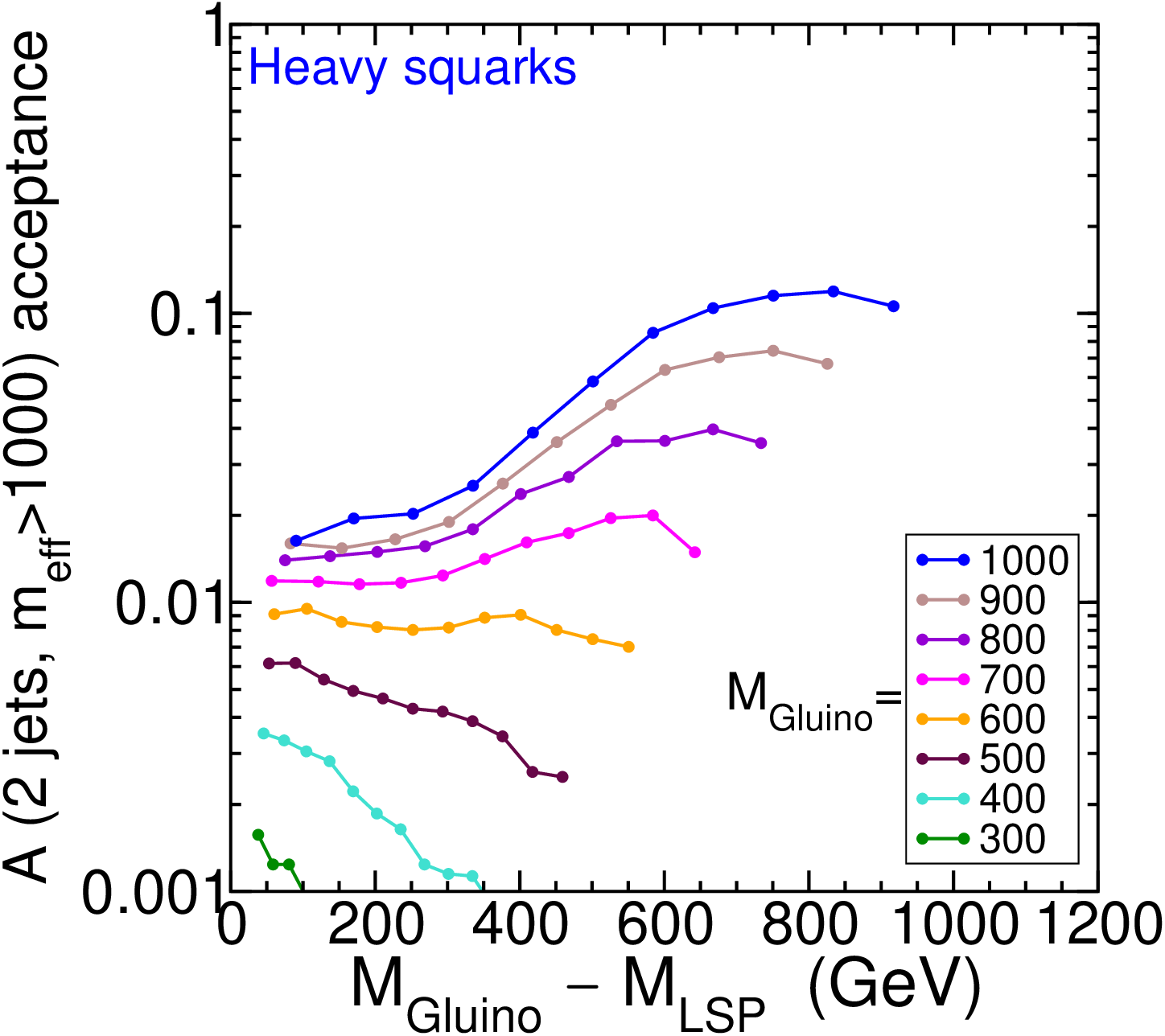}
\includegraphics[width=8.3cm,angle=0]{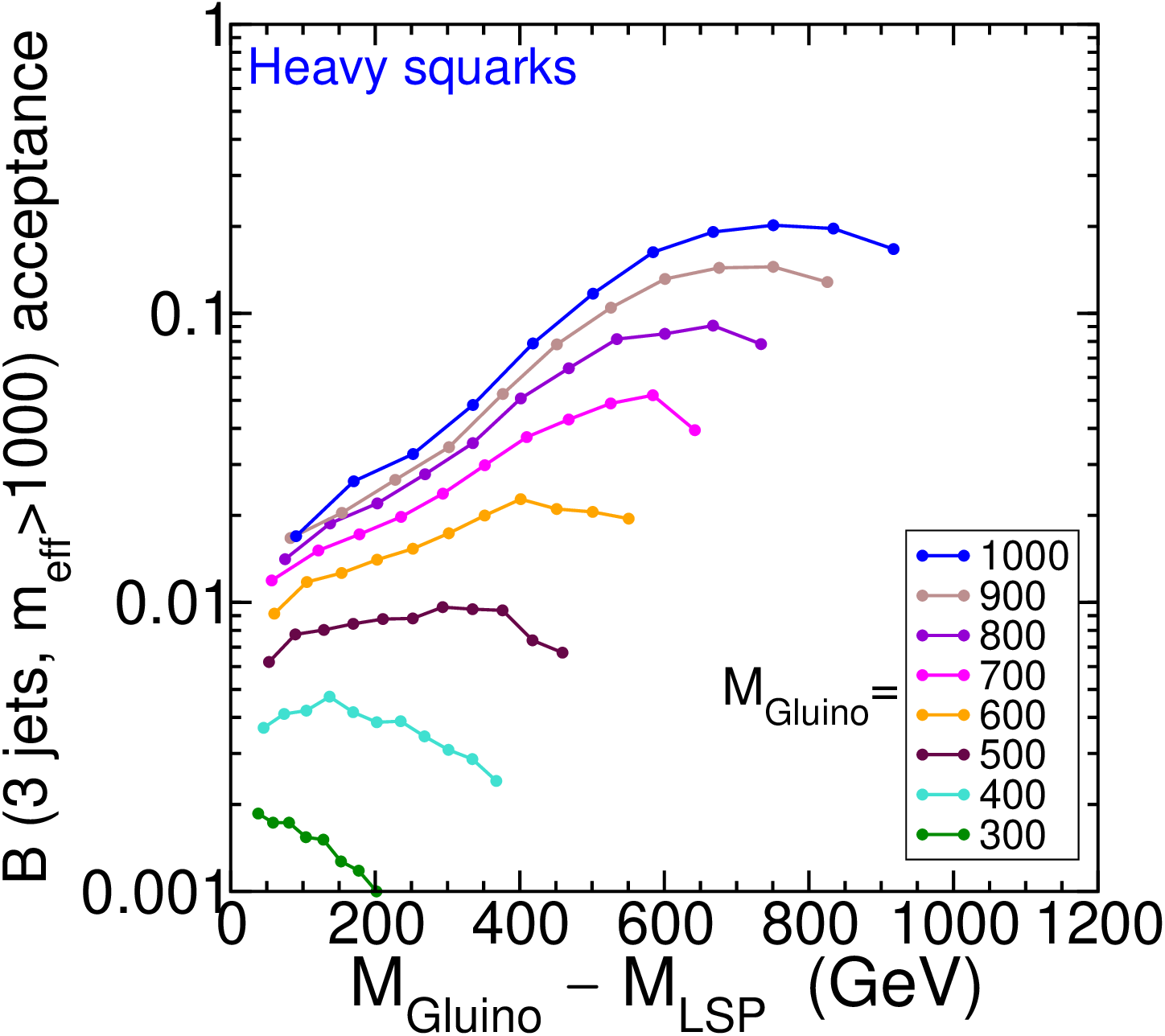}}

\vspace{0.2in}

\mbox{\includegraphics[width=8.3cm,angle=0]{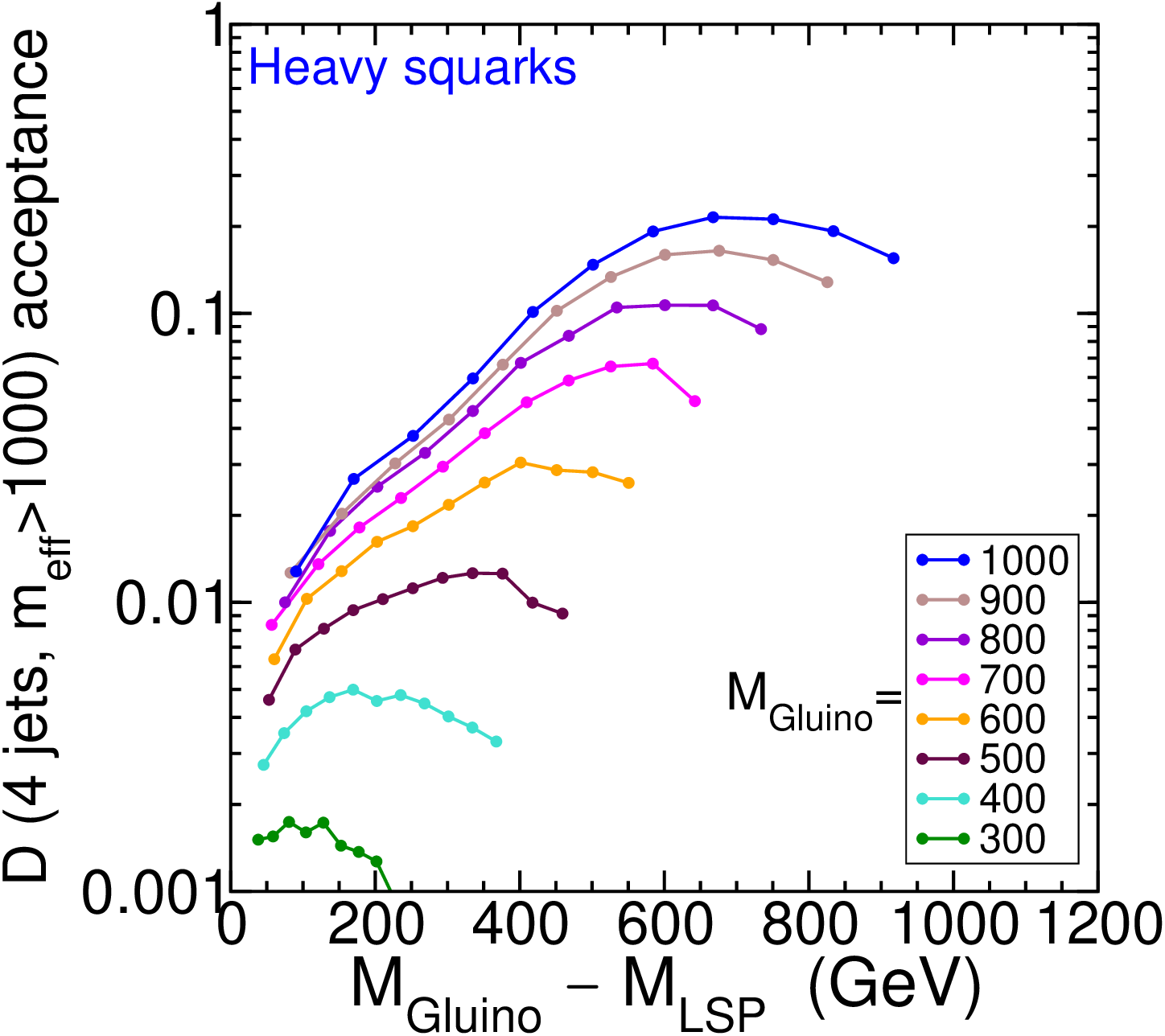}
\includegraphics[width=8.3cm,angle=0]{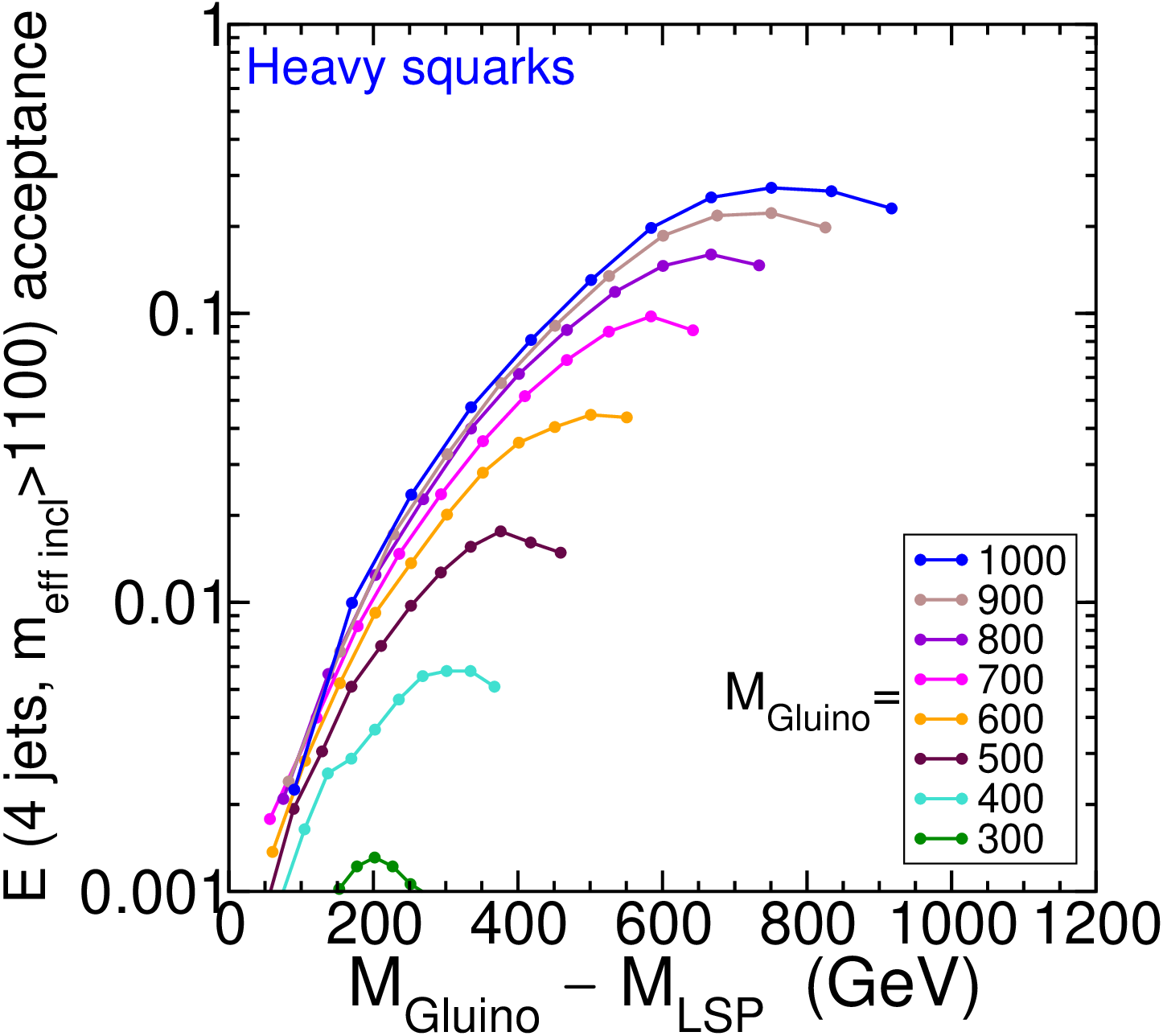}}
\caption{\label{fig:accHSQ}
As in Figure~\ref{fig:acc}, but for the models II (heavy squarks) 
defined in the text.}
\end{figure}

The acceptances for the heavy squark models of class II are similarly 
shown in Figure \ref{fig:accHSQ} for gluino masses from 300 up to 1000 
GeV. Here the acceptance tends to increase more steadily with increasing 
$M_{\tilde g}$. There is sometimes a notable 
decrease in 
acceptance for signal regions A and B as the mass difference $M_{\tilde 
g} - M_{\rm LSP}$ increases, so that the 
largest acceptances are achieved with non-zero compression, that is, when 
$M_{\tilde g} - M_{\rm LSP}$ is not maximum. This perhaps 
surprising effect, studied in 
\cite{LeCompte:2011cn}, is due to the fact that as the compression 
increases, the $\meff$ distribution becomes soft faster than the 
$\missET$ distribution does, so that more events pass the $\missET/\meff$ 
cuts.

The acceptances for models in classes III (heavy winos) and IV (heavy squarks and winos) are qualitatively similar, and so are not shown.

In Figure \ref{fig:sigacc}, we show contours of cross-section times 
acceptance for all five signal regions, for each of the model classes I, 
II, III, IV in separate panels. The contours for each signal 
region are for the corresponding ATLAS 
limits on 
non-Standard Model processes listed in Table \ref{tab:cuts} (taken from 
\cite{ATLASsummer2011}), so that the regions to the left of each contour 
may be regarded as the approximate exclusion regions for that signal 
definition. In the panels for model classes I and 
II, the (orange) dotted line indicates the case $c=0$ in 
which the 
ratio of 
gaugino masses at the TeV scale is approximately the same as mSUGRA. For 
the light squark class of models I, 
the best signal regions for exclusion are A (when $M_{\tilde g} - M_{\rm 
LSP} > 400$ GeV) and B (for most of the range of smaller mass 
differences, except for the most extremely compressed case).

\begin{figure}[!tbp]
\mbox{\includegraphics[width=8.3cm,angle=0]{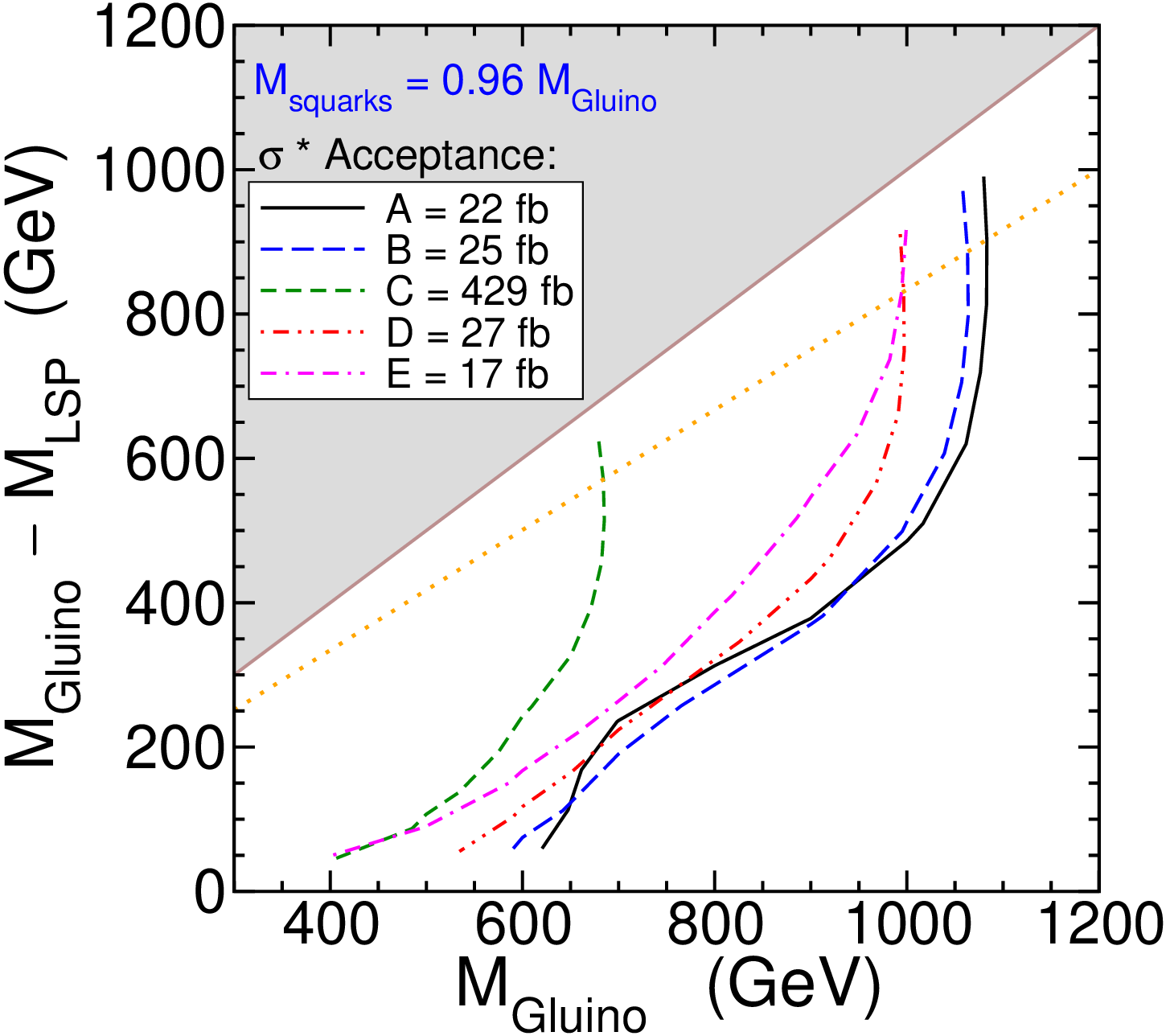}
\includegraphics[width=8.3cm,angle=0]{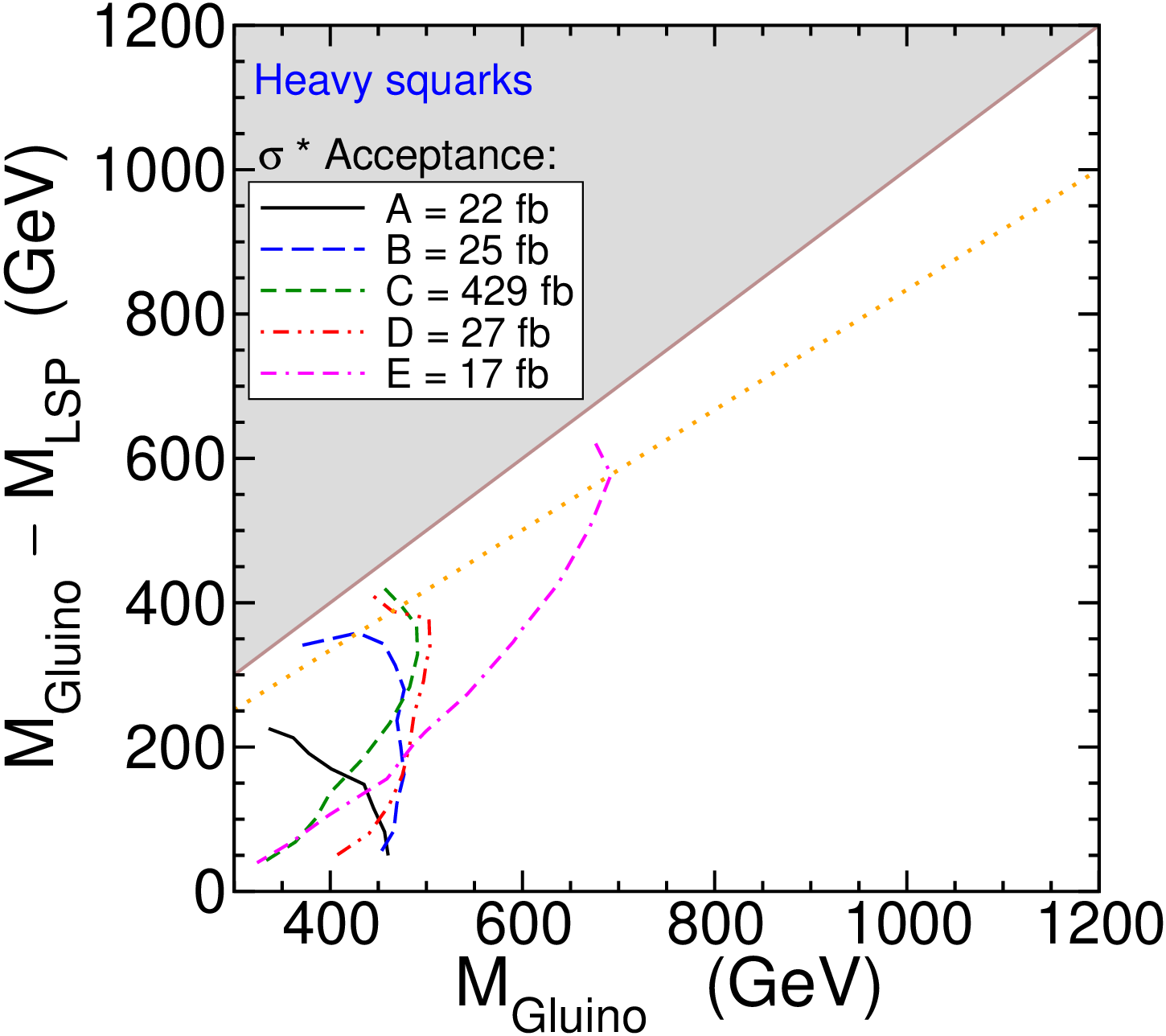}}

\vspace{0.2in}

\mbox{\includegraphics[width=8.3cm,angle=0]{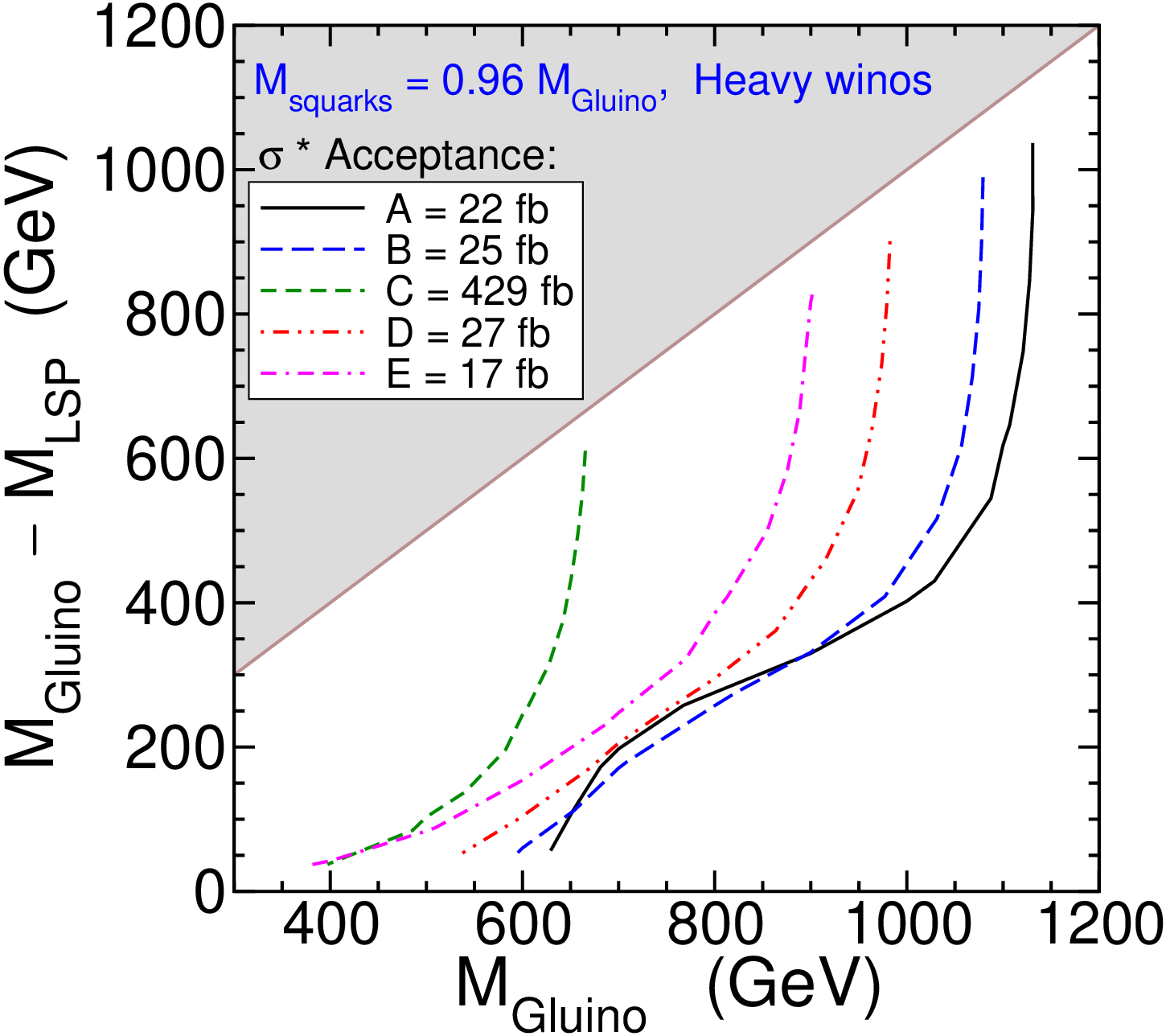}
\includegraphics[width=8.3cm,angle=0]{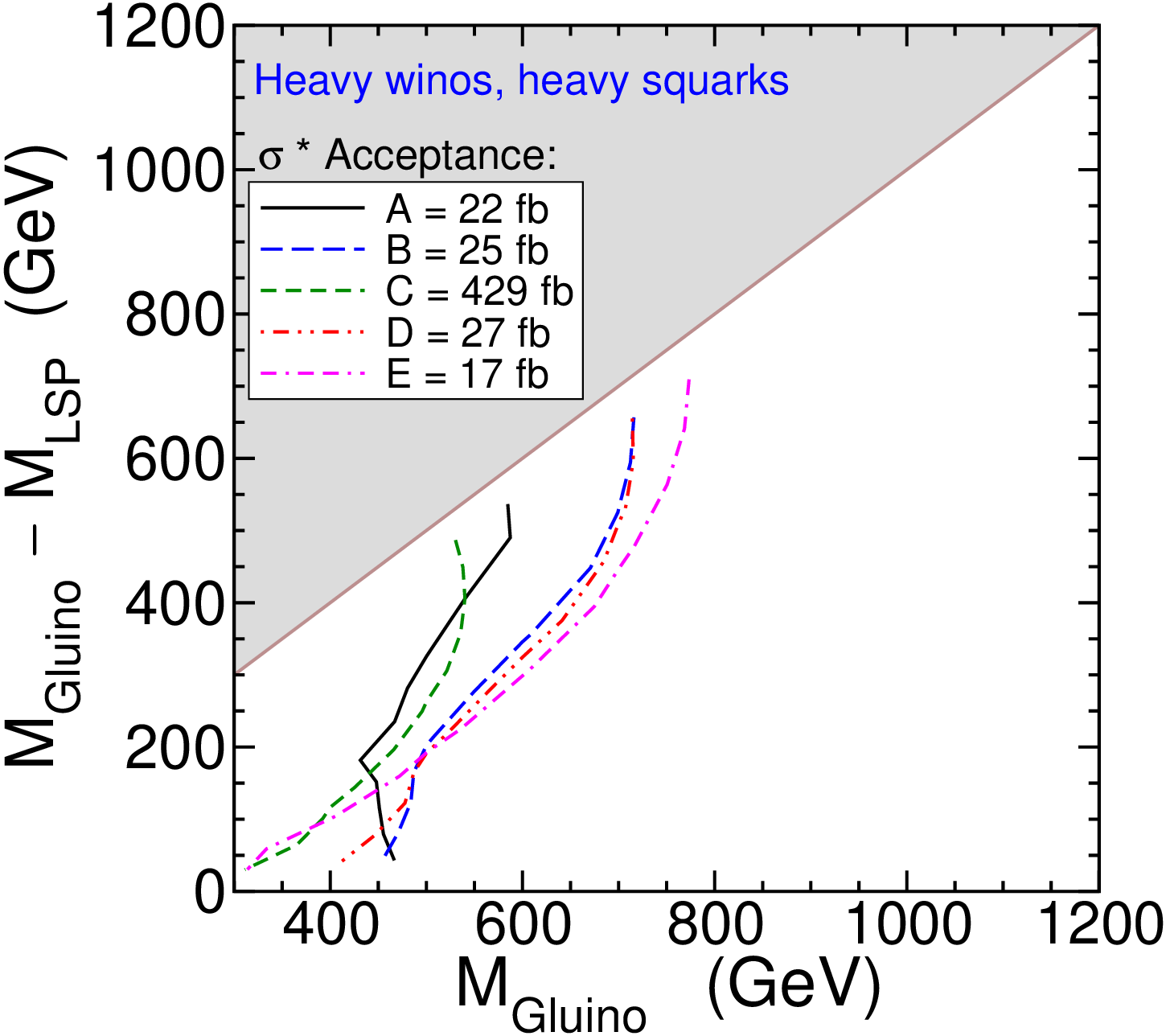}}
\caption{\label{fig:sigacc}
Contours of constant cross-section times acceptance 
for the five signal regions defined in Table \ref{tab:cuts},
in the $M_{\tilde g} - M_{{\tilde N_1}}$ vs. $M_{\tilde g}$ plane 
obtained by varying the gaugino mass compression parameter 
$c$ between $-0.1$ and $0.9$. The four panels correspond to the 
models I (light squarks), II (heavy squarks), III (heavy winos),
and IV (heavy winos and heavy squarks). 
The dotted lines in the first two cases
corresponds to the mSUGRA-like case $c=0$. 
Increased compression is lower in each plane.}
\end{figure}
Signal regions A and B likewise give the best exclusions for models in 
class III (heavy winos but light squarks). Because the gluino and squark 
decays in this class of models do not pass through the intermediate 
cascade step of winos, the visible energy per jet tends to be larger, 
leading to stronger exclusions as shown. In both of the model classes I 
and III with squarks slightly lighter than the gluino, we find that even 
in the case of extreme compression one can still set a limit of better 
than $M_{\tilde g} > 600$ GeV.

In models with heavy squarks, the limits are much worse, since the main 
SUSY production is only gluino pairs. For models of class II with heavy 
squarks and light winos, the best limit is set using the high-mass signal 
region E when $M_{\tilde g} - M_{\rm LSP} > 200$ GeV. For smaller mass 
differences, the 3-jet signal region B sets better limits, because the 
$\meff$ distribution for the signal events becomes too soft. Still, it 
should be possible to set a limit of about $M_{\tilde g} > 450$ GeV 
using signal regions A and B, even 
in the case of extreme compression with $c=0.9$. Qualitatively similar 
statements apply to models in class IV with both squarks and winos 
decoupled. Note than in all cases, signal region C is comparatively 
ineffective in setting limits, because the backgrounds are too large.

\clearpage

\begin{figure}[!tbp]
\mbox{\includegraphics[width=8.3cm,angle=0]{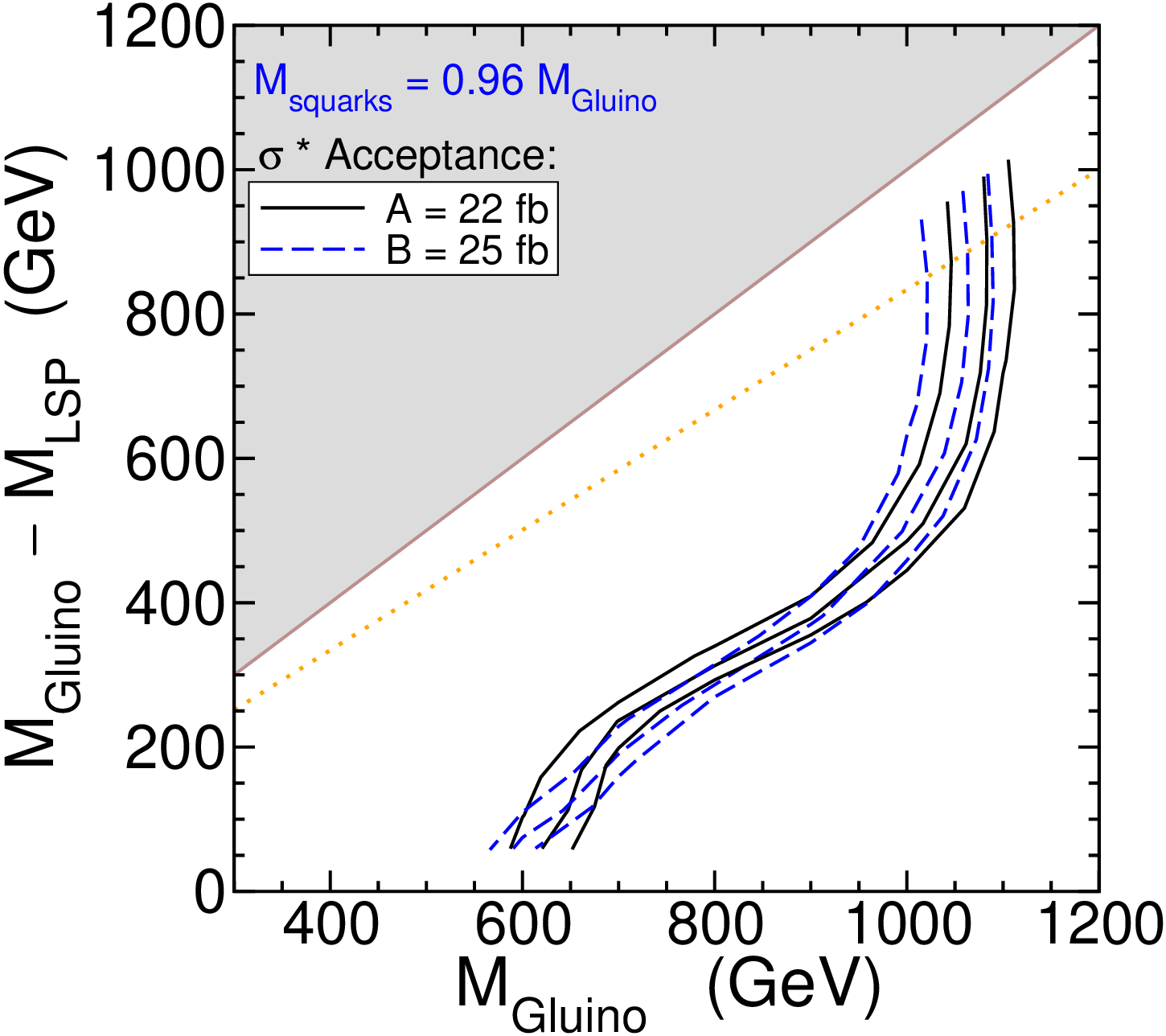}
\includegraphics[width=8.3cm,angle=0]{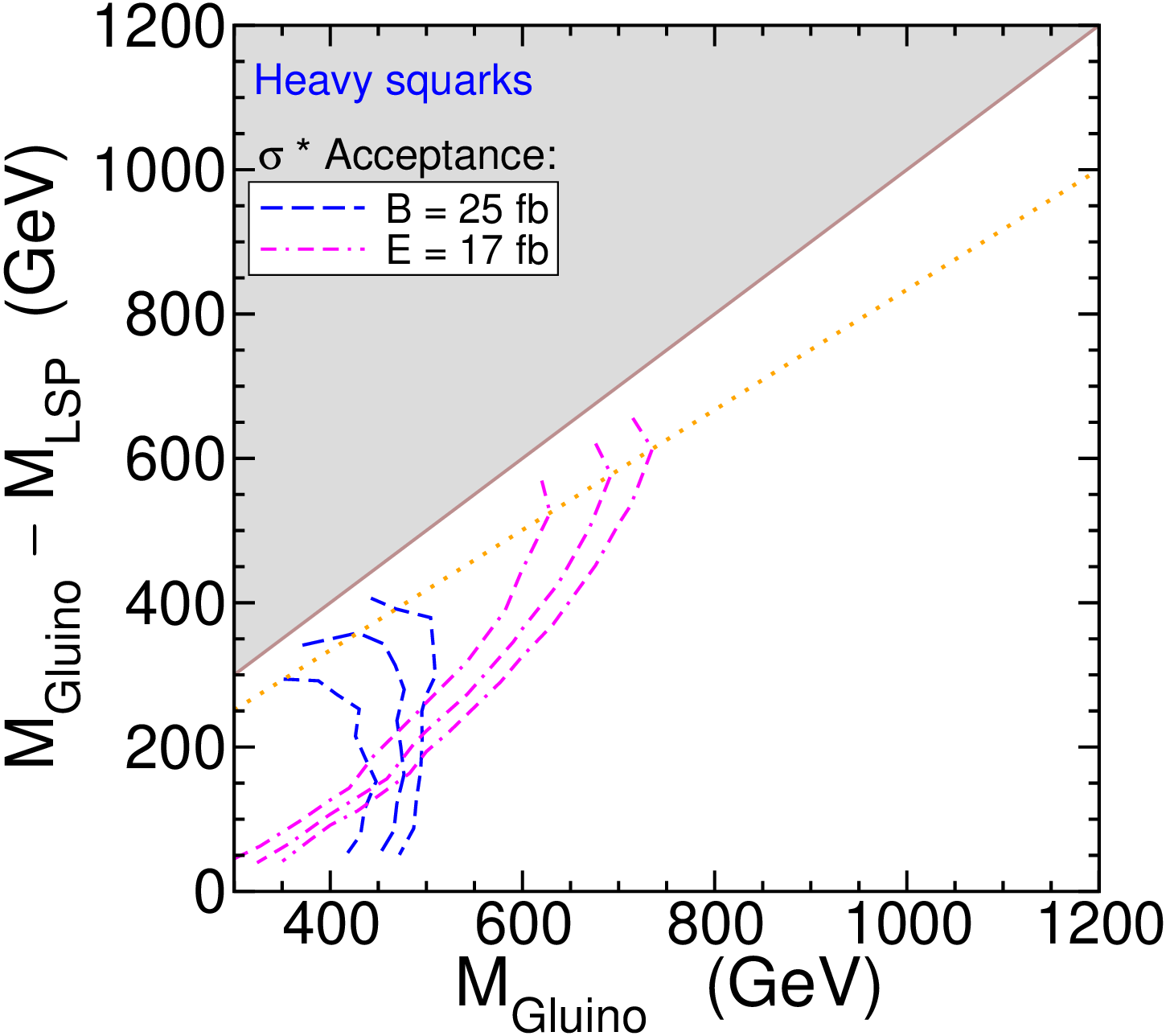}}
\caption{\label{fig:systematic}
The impact of systematic uncertainties in the signal rates on contours of 
constant cross-section times acceptance 
for selected signal regions. The middle line in each case is the same 
result as in Figure \ref{fig:sigacc}. The corresponding left and right 
contours 
show the impact of decreasing and increasing the total signal 
production rate by 25\%. The two panels correspond 
to the models I (light squarks, left) and II (heavy squarks, right).}
\end{figure}

The preceding results were all obtained using the Prospino NLO 
default renormalization and factorization scale choices and without 
taking into account the possible systematic uncertainties in the signal 
production cross sections. In general, uncertainties in QCD production cross-sections
are notoriously difficult 
to estimate. It is well-known that variation of 
renormalization and factorization scales and PDFs do not give reliable 
estimates of the production cross-section uncertainties. To illustrate 
the 
potential impacts of these uncertainties, we show in Figure 
\ref{fig:systematic} how the results vary when changing the assumed total 
signal cross-section by $\pm 25$\% for the models of class I and II. Only 
the two signal regions that give the strongest limits over the most 
significant ranges of $M_{\tilde g} - M_{\rm LSP}$ are shown in each 
case. This variation results in a change in the gluino mass limit in 
these models that can exceed $\pm$50 GeV, depending on the model 
parameters.


\section{Outlook}
\label{sec:conclusion}
\setcounter{equation}{0}
\setcounter{footnote}{1}

In this paper, we have studied the reach of 1 fb$^{-1}$ of LHC data
at $\sqrt{s} = 7$ TeV for compressed SUSY models, extending our earlier 
results for 35 pb$^{-1}$ in \cite{LeCompte:2011cn}. We found that even in 
the most compressed case studied, in which the gluino is only about 9\%
heavier than the LSP, the limit on the gluino mass should be about
$M_{\tilde g} > 600$ GeV for squarks that are slightly lighter than the 
gluino, and about $M_{\tilde g} > 450$ GeV when squarks are very heavy.
The best limits (and discovery potential) come from signal regions which 
require 2 or 3 jets. In designing future searches for compressed 
SUSY, it is probable that the best reach will be obtained by increasing 
the cut on $\missET$ as necessary to reduce the backgrounds, rather 
than by very hard cuts on $\meff$ (or $H_T$). 
This is 
because as the compression increases,
both the $\missET$ and $\meff$ distributions get softer, but the latter 
more drastically. 
(A more precise quantitative statement about this is beyond the scope of 
this paper, since it would require detailed background estimates 
including crucially detector response-specific information.) 
Future searches should take into account that signal 
regions optimized for mSUGRA and for simplified models with massless or 
light LSPs will therefore not do very well for compressed SUSY models, 
and this effect will become more significant as higher mass scales are 
probed.

{\em Acknowledgments:} 
The work of TJL was supported in part by the U.S. Department of Energy, 
Division of High Energy Physics, under Contract DE-AC02-06CH11357.
The work of SPM was supported in part by the 
National Science Foundation grant number PHY-1068369. SPM is grateful
for the hospitality and support of the Kavli Institute for Theoretical 
Physics in Santa Barbara.
This research was supported in part by the National Science Foundation 
under Grant No. PHY05-51164.


\end{document}